\documentclass[accepted,twoside]{article}
\usepackage[utf8]{inputenc}
\usepackage[T1]{fontenc}
\usepackage{xspace}
\usepackage{graphicx}
\usepackage{grffile}
\usepackage{longtable}
\usepackage{wrapfig}
\usepackage{rotating}
\usepackage[normalem]{ulem}
\usepackage{amsmath}
\usepackage{textcomp}
\usepackage{amssymb}
\usepackage{capt-of}
\usepackage{aistats2020}
\usepackage{amsmath}
\usepackage{subcaption}
\usepackage{hyperref}
\usepackage[capitalise]{cleveref}
\usepackage{todonotes}
\usepackage{natbib}
\author{Alejandro Catalina}
\date{\today}
\title{Projection predictive inference beyond Exponential Family models}
\hypersetup{
  pdfauthor={Alejandro Catalina},
  pdftitle={Projection predictive inference beyond Exponential Family models},
  pdfkeywords={projpred, exponential family, kullback-leibler},
  pdfsubject={Extend projective inference beyond exponential family models},
  pdfcreator={Emacs 26.3 (Org mode 9.4)},
  pdflang={English}}
  
\newcommand{\kakapo}{k\={a}k\={a}p\={o}\xspace}
\begin{document}

\twocolumn[
\aistatstitle{Latent Space Projection Predictive Inference}
\aistatsauthor{Alejandro Catalina$^1$ \And  Paul B\"urkner$^2$ \And Aki Vehtari$^1$}
\aistatsaddress{$^1$ Helsinki Institue for Information Technology, HIIT, Aalto University, Finland \\ $^2$ Cluster of Excellence SimTech, University of Stuttgart, Germany}
]

\begin{abstract}
Given a reference model that includes all the available variables, projection predictive inference replaces its posterior with a constrained projection including only a subset of all variables.
We extend projection predictive inference to 
enable computationally efficient variable and structure selection in models outside the exponential family.
By adopting a latent space projection predictive perspective we are able to: 
1) propose a unified and general framework to do variable selection in complex models while fully honouring the original model structure, 
2) properly identify relevant structure and retain posterior uncertainties from the original model, and
3) provide an improved approach also for non-Gaussian models in the exponential family.
We demonstrate the superior performance of our approach by thoroughly testing and comparing it against popular variable selection approaches in a wide range of settings, including realistic data sets.
Our results show that our approach successfully recovers relevant terms and model structure in complex models, selecting less variables than competing approaches for realistic datasets.
\end{abstract}

\setcitestyle{round}

\section{Introduction}
\label{sec:org806514f}

Variable and structure selection plays an important role in a robust Bayesian workflow~\citep{gelman2020bayesian}.
While variable selection has been extensively studied and successfully applied for models in the exponential family~\citep{Koopman_1936}, it has not received the same attention regarding models outside of the exponential family (e.g., advanced ordinal, count or time-to-event (also known as survival) data distributions), despite their important applications in different fields~\citep{kelter_survival,nagler_ordinal,burkner2019ordinal,barron1992analysis}.

We propose an efficient, stable, and information theoretically justified method to make variable selection for non-normal observation models in or beyond the exponential family.
The main benefits of the proposed \textit{latent space projection predictive inference} are: 
\begin{enumerate}
\item we enable the projection predictive variable and structure selection for models outside the exponential family while honouring the original model structure and its predictive uncertainty, 
\item we obtain more stable projections for non-Gaussian exponential family models, 
\item we demonstrate the superior performance of our method as compared to state-of-the-art competitors in both simulated and real-world scenarios, 
\item we provide a ready-to-use open source implementation of the new methods.
\end{enumerate}

\section{Projection predictive inference}
\label{KL-div-proj}

Given posterior draws $\{\lambda_{*}^{(s)}\}_{s=1}^S \sim p(\lambda \mid \mathcal{D})$ from a reference model with data $\mathcal{D} = \{X, y\}$, outcome observations $y$ and predictor variables $X$, projection predictive inference~\citep{piironen_projective_2018,catalina2020projection} learns a projection $q_\bot(\lambda)$ containing only a subset of variables that matches the reference predictive performance as close as possible.
The solution is given by the minimiser of the Kullback-Leibler (KL) divergence from the reference model to the projection predictive distributions~\citep{Dupuis2003}.
Let $p(\tilde{y} \mid {\cal D})$ be the reference predictive distribution and $\{\lambda_{\bot}^{(s)}\}_{s=1}^S \sim q_\bot(\lambda)$ be draws from the projection, then:
\begin{align}
  \text{KL} &\left( p \left( \tilde{y} \mid {\cal D} \right) \parallel q_\bot \left( \tilde{y} \right) \right) \nonumber \\
            & = - \mathbb{E}_{\lambda_{*}} \left( \mathbb{E}_{\tilde{y} \mid \lambda_*} \left( \log \mathbb{E}_{\lambda_\bot} \left( p \left( \tilde{y} \mid \lambda_\bot \right) \right) \right) \right) + \text{C}. 
  \label{eq:kl_proj}
\end{align}

As the integrals involved are in most cases computationaly infeasible, \cite{Goutis_1998} suggest to change the order of integration and minimisation and solve the optimisation for each posterior draw separately 
\begin{equation}
    \text{arg}\max_{\lambda}\mathbb{E}_{\tilde{y} \mid \lambda_*^{(s)}} \left( \log \mathbb{E}_{\lambda_\bot} \left( p \left( \tilde{y} \mid \lambda \right) \right) \right).
    \label{eq:KL-by-draw}
\end{equation}

\cite{piironen_projective_2018} proposed a further speed-up by first clustering the posterior draws $\lambda_*^{(s)}$ and then solving the optimisation individually for each resulting cluster centre $\{\lambda_*^{(c)}\}_{c=1}^C$.

For models in the exponential family distribution, \cref{eq:KL-by-draw} coincides with computing maximum likelihood estimates under the projection model as
\begin{equation}
  \lambda_{\bot} = \arg \max_{\lambda} \sum_{i=1}^N\mu_{i}^{*}\xi_i(\lambda) - B(\xi_i(\lambda)),
  \label{eq:maximum_likelihood_estimates}
\end{equation}
where \(\mu^{*} = \mathbb{E}_{\tilde{y} \mid \lambda_*}(\tilde{y})\) are mean predictions of the reference model, $\xi_i$ are the natural parameters for the $i$th observation and $B(\cdot)$ is a function of the natural parameters.
These maximum likelihood estimates can be efficiently computed by solving penalised iteratively reweighted least squares \citep[PIRLS;][]{marx1996iteratively}. 
If non-constant, the projected scale parameter of the exponential family model  is then obtained as
\begin{equation}
   \phi_\bot = \text{arg}\max_\phi\sum_{i=1}^N\left(\dfrac{r_i(\lambda_\bot)}{A(\phi)} + \mathrm{E}_{\tilde{y}_i\mid\lambda_*}(H(\tilde{y}_i, \phi)) \right),
   \label{eq:dispersion}
\end{equation}
where $A, H$ are family-specific functions and $r_i(\lambda_\bot) = \mu_i^*\xi_i(\lambda_\bot) - B(\xi_i(\lambda_\bot))$ does not depend on $\phi$. 

\section{Latent space projective inference}
\label{sec:latent}

\begin{figure*}[tp]
  \centering
  \begin{subfigure}{0.45\linewidth}
    \includegraphics[width=\textwidth]{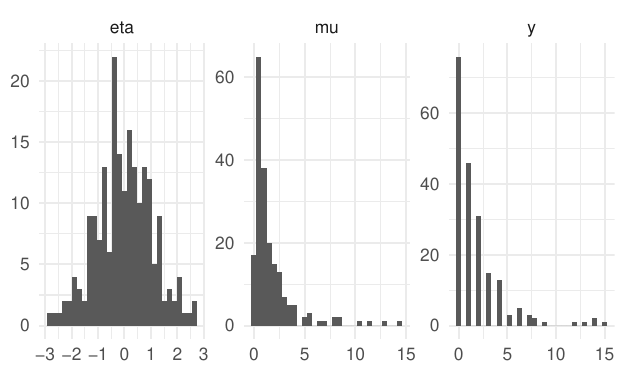}
    \caption{Latent linear predictor, transformed predictor and response space for a Poisson model with \texttt{log} link function.}
    \label{fig:illustration_pois}
  \end{subfigure}
  ~~~~
  \begin{subfigure}{0.45\linewidth}
    \includegraphics[width=\textwidth]{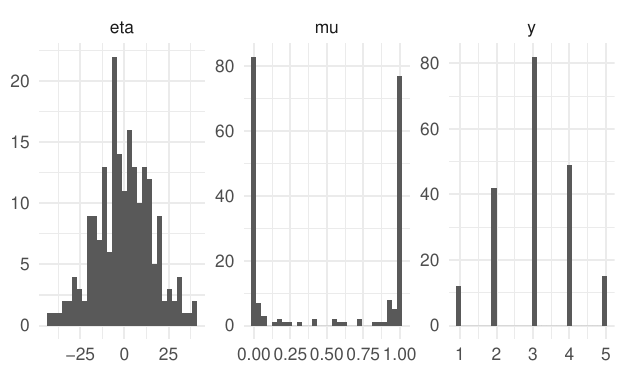}
    \caption{Latent linear predictor, transformed predictor and response space for an Ordinal cumulative model with \texttt{probit} link function.}
    \label{fig:illustration_cumulative}
  \end{subfigure}
  \caption{Latent, transformed and response space representation for different models.}
  \label{fig:eta-vs-mu}
\end{figure*}

The equivalence between maximum likelihood estimates and KL minimiser does not hold for models outside the exponential family.
For data ${\cal D} = \{X, y\}$, parameters $\lambda$ and inverse link function $g$, we assume the general model formulation
\begin{equation}
  y \sim p \left( \mu, \phi \right),\quad \mu  = g(\eta),\quad \eta \sim p \left( \lambda \mid X \right),
  \label{eq:model_formulation}
\end{equation}
where $\eta$ is commonly called the latent predictor~\citep{McCullagh_1989}. 
In~\cref{fig:eta-vs-mu} we show the difference between the linear and the transformed predictor spaces for two common models, a Poisson model with a \texttt{log} link function and an ordinal cumulative model~\citep{burkner2019ordinal} with \texttt{logit} link function.

We propose to reformulate the projection problem by solving the KL minimisation in the latent predictive space $p(\tilde{\eta} \mid \lambda_*)$ as:
\begin{align*}
  \text{KL} & \left( p \left( \tilde{\eta} \mid {\cal D} \right) \parallel q_{\bot} \left( \tilde{\eta} \right) \right) = \\
            &  - \mathbb{E}_{\lambda_{*}} \left( \mathbb{E}_{\tilde{\eta}\mid\lambda_*} \left( \log \mathbb{E}_{\lambda_\bot} \left( p \left( \tilde{\eta} \mid \lambda_\bot \right) \right) \right) \right) + \text{C}. \nonumber
\end{align*}

\subsection{Non exponential family case}
The distribution $p(\eta\mid\lambda)$ is model-dependent and in non-exponential family case doesn't, in general, have a nice closed form. 
We propose to approximate $p(\eta\mid\lambda)$ with a Gaussian distribution, a choice motivated by several reasons: 
1) as potential boundaries on $\mu$ are enforced only via the inverse link $g$, $\eta$ itself is unbounded so that its support matches the support of a Gaussian, 
2) the Gaussian distribution belongs to the exponential family and is thus computationally easy to handle, and
3) the model on $\eta$ is typically additive, such that commonly chosen Gaussian priors on the latent parameters $\lambda$ make the implied distribution of $\eta$ closer to Gaussian as well. 
Now that the transformed projection is again in the exponential family, we can apply the approach outlined in Section \ref{KL-div-proj} to solve the projection. 

Although the dispersion of the latent space is typically not a model parameter, we still need to approximate it to compute its projection in~\cref{eq:dispersion}. 
The approximate dispersion is model dependent, and may sometimes even be known analytically. 

\subsection{Exponential family case}

The latent projection formulation also works for models in the exponential family. 
For Gaussian observation models, it coincides exactly with the original framework, and therefore no improvement is gained with the latent approach. 

For non-Gaussian exponential family models, the original framework computes an approximate solution to~\cref{eq:kl_proj} via PIRLS (as in~\cref{eq:maximum_likelihood_estimates}), which often results in unstable solutions for models with complex structure or link functions~\citep{catalina2020projection}.
In contrast, the latent approach computes an approximation on the latent space, removing the complexity of the link function and the response model.
Experimental results show that the latent approach results in significant improvements also for non-Gaussian exponential family models.

\section{Related work}
\label{sec:org2d23ce0}

For models in the exponential family, variable selection has a strong presence in the literature.
Some methods perform variable selection by optimising a penalised likelihood formulation~\citep{tibshirani_lasso_1996,Zou_2005,Friedman_2010,Candes_2007,breiman_garrote_1995,fan_nonconcave_2001}, while at the same time trying to select a subset of relevant variables.
These methods impose a penalisation on large coefficients, effectively driving some of them towards zero, depending on the choice of regularization.
For more details, we refer the reader to~\cite{Hastie_2015}.
These methods suffer from several drawbacks.
First, by dealing with the estimation of the model and the selection of variables at the same time, they often result in suboptimal solutions~\citep{piironen_comparison_2017}.
Second, they are derived for specific likelihoods, and generalising them to other models is difficult, if at all possible.
Most notably, these methods cannot perform variable selection on group-specific parameters in hierarchical models~\citep{catalina2020projection}.

These approaches have been generalized to specific likelihoods outside of the exponential family, such as ordinal or Cox models~\citep{wurm17:_regul_ordin_regres_r_packag,glmnetcr,ordinalgmifs,fan2005overview}, but no general framework exists.
Penalised likelihood approaches can be applied to models outside of the exponential family by approximating the likelihood with an exponential family distribution.
This enables variable selection but the resulting model is likely to underperform in terms of predictive performance.
A model in the original response space can be obtained by fitting a Bayesian model including only the selected variables.
We compare our method against baselines that follow this approach in~\cref{sec:synth-exper}.

Variable selection is also addressed from the Bayesian perspective~\citep{ohara2009review}.
This is typically done by imposing so called sparsity priors, such as the horseshoe~\citep{Carvalho_2010,piironen_sparsity_2017} or the spike-and-slab~\citep{Ishwaran_2005}.
As the posterior itself is not fully sparse, a sparse solution is then obtained by thresholding based on posterior expectations. 

If a Markov chain Monte Carlo (MCMC) algorithm~\citep{robert2013monte} is used for inference, no strong assumptions on the likelihood or overall structure are needed.
This makes Bayesian inference applicable to models outside of the exponential family with multilevel (or other complex) structure.
This helps with the generality of the inference, but still falls short on the selection, since the user still needs to manually decide which variables should be selected.

Bayesian reference models have been used for variable and structure selection tasks in the context of exponential family models, including (additive) multilevel models~\citep{piironen_projective_2018,catalina2020projection,pavone_2019,piironen_projection_2016}.
However, its application to models outside of the exponential family has remained unexplored.

Our approach solves variable selection on non exponential family models by performing the selection on the latent space of the original model, therefore keeping the original structure.
We approximate the unknown latent distribution with a Gaussian. 
The problem of learning an implicit distribution has been approached in the Bayesian inference literature from different angles.
Some authors tackle this problem with variational inference~\citep{blei2017variational}, by learning an implicit mapping between samples of the implicit distribution and a powerful and expressive learner, typically normalising flows~\citep{rezende2016variational,DBLP:journals/corr/Huszar17,pequignot2020implicit,titsias2019unbiased}.
Optimising these flexible models is typically expensive, requiring many iterations and diagnostics to have any guarantee of convergence~\citep{DBLP:conf/nips/DhakaCAMHV20}.
On top of that, the main bottleneck in projection predictive inference is not solving this projection \emph{once}, but possibly many times, as complex models require many posterior draws to be projected for the projections to fully capture the posterior uncertainty in the reference model.
\begin{figure*}[tp]
  \centering
  \begin{subfigure}[tp]{0.45\linewidth}
    \includegraphics[width=\linewidth]{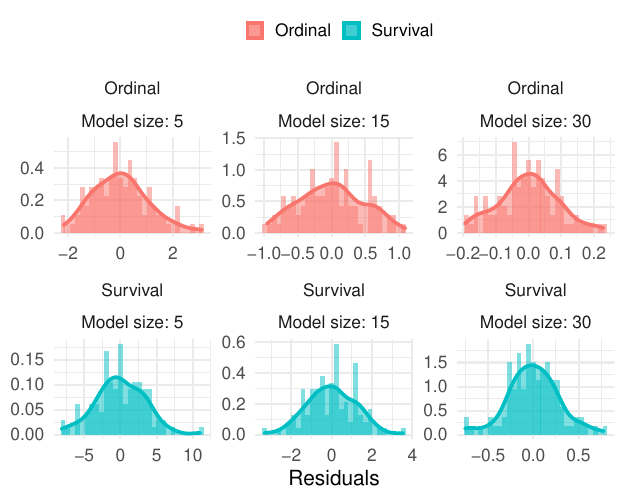}
    \caption{Residuals histogram plots for three selected projections for a reference model of size $D = 50$. In all cases, the residuals distribution is close to normal with a decreasing standard deviation as more terms are introduced in the projection. Notice the shrinking $x$-axis as the projections include more terms.}
    \label[subfigure]{fig:residuals_combined}
  \end{subfigure}
  ~~~~
  \begin{subfigure}[tp]{0.45\linewidth}
    \includegraphics[width=\linewidth]{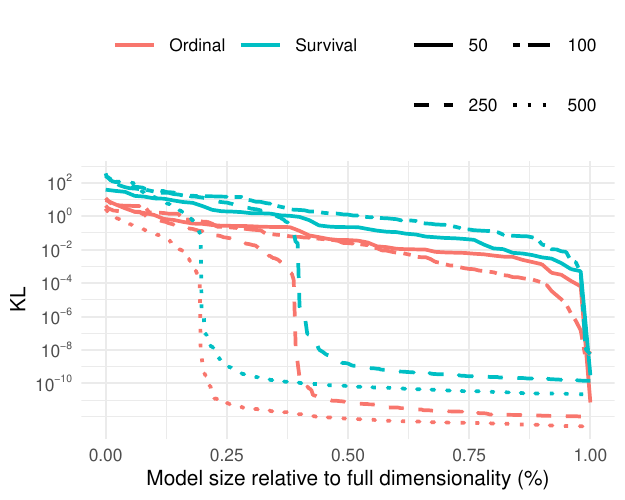}
    \caption{KL divergence between reference and projection predictive distributions. Different linetypes indicate size of the reference model. As projections include a higher proportion of the reference model, the KL-divergence approaches 0 faster.}
    \label[subfigure]{fig:kl_combined}
  \end{subfigure}
  \caption{Residuals histogram and KL-divergence plots for projections on a simulated time-to-event survival analysis model and an ordinal cumulative model, both with $N = 100$ observations, and correlation factor $\rho = 0.3$.}
  \label{fig:kl_residuals}
\end{figure*}

\section{Experiments}
\label{sec:org94584cd}

We evaluate the performance of the proposed method with simulated and real-world data experiments.
In particular, we must account for complex high dimensional posterior geometries, the role of correlated terms, and the size of the selected subset of variables in different models outside the exponential family.

First, we demonstrate the use of different diagnostics for the latent approximation.
Second, we demonstrate the variable selection itself in high dimensional non-exponential family examples.
Then, we extend the experiments to real datasets.
Finally, we show the benefits that our method brings to models in the exponential family, too.

For the implementation of our new approach, we use modified \texttt{projpred}~\citep{projpred_package}.
The implementation of the latent projection predictive inference approach is publicly available at \url{https://github.com/stan-dev/projpred/tree/latent_projection}.

\subsection{Diagnosing the quality of the latent approximation}
\label{sec:residuals-experiments}
We use two diagnostics to asses the quality of the approximate projections:
\begin{itemize}
  \item We Perform projection predictive checks on the residuals $\tilde{\eta}_{*} - \tilde{\eta}_{\bot}$, where $\tilde{\eta}_{*}$ corresponds to the latent predictions of the reference model and $\tilde{\eta}_{\bot}$ to the predictions of the projection.
  \item We check the Kullback-Leibler divergence between the reference's and projection's predictions after convergence, which should approach $0$ as more terms are included in the projection.
\end{itemize}

We use simulated data from 1) an ordinal cumulative model with \texttt{probit} link function and 2) a time-to-event survival model with a \texttt{log} link function.
The outcome in both cases is generated as a function of $D \in [50, 100, 250, 500]$ sampled predictors with a uniform correlation factor of $0.3$, where only $60$\% of the predictors have a non-zero effect on the response.
The full generation process for these data is detailed and further analysed in~\cref{sec:synth-exper}.

\paragraph{Projection predictive checks assess normality assumptions.}
Histograms of projection residuals (\cref{fig:residuals_combined}) for various model sizes are a practical diagnostic for the normality assumption.
As more terms enter the projection, the residuals get smaller and more concentrated, as indicated by the shrinking $x$-axis in the figure.

\paragraph{Kullback-Leibler divergence shows that the latent projections eventually match the reference model predictions.}
Even for the most challenging scenarios, the KL-divergence of the latent predictions shows that the projection predictive distribution gets closer (the KL-divergence approaches $0$) to the reference predictive distribution as more terms enter the projection (\cref{fig:kl_combined}).

\subsection{Non-exponential family models with simulated data}
\label{sec:synth-exper}

We compare the predictive performance of the optimal submodels in terms of held-out expected log predictive density (ELPD)~\citep{aki_model_assessment} for two types of models: an ordinal cumulative model and a time-to-event survival analysis model with a Weibull hazard process.
Additionally, we examine the performance regarding the selection of truly relevant variables.

For the simulated high-dimensional data, we compare the performance of our approach to other popular sparsifying solutions in the literature:
\begin{itemize}
  \item Elastic net regularization as implemented by \texttt{glmnet}~\citep{glmnet}, abbreviated as \texttt{glmnet} in the figures,
  \item Spike-and-slab sparsifying priors as implemented by \texttt{spikeSlabGAM}~\citep{spikeslabgam}, abbreviated as \texttt{ss} in the figures,
  \item Spike-and-slab LASSO priors as implemented by \texttt{SSLASSO}~\citep{sslasso}, abbreviated as \texttt{sslasso},
  \item Projection predictive inference on the approximate response space as implemented by \texttt{projpred}~\citep{projpred_package}, abbreviated as \texttt{projpred} in the figures.
\end{itemize}

Since these approaches do not exist for the specific likelihoods we use, or for models outside the exponential family in general, we run variable selection on an approximate model, where we assume a normal likelihood of the response rather than the appropriate likelihood (cumulative or Weibull in our examples).
Note that this normal approximation is different from our latent approach, since the latter approximates the latent predictor, not the response.

\begin{figure}[tp]
  \centering
  \includegraphics[width=\linewidth]{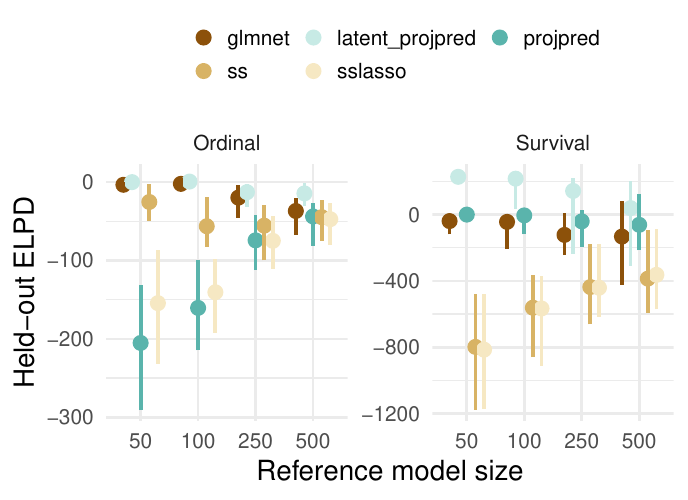}
  \caption{Median and 90\% confidence interval ELPD difference to the reference model for the selected subset of predictors for an ordinal cumulative model (left) and a time-to-event survival model (right) with $N = 100, \rho = 0$. Higher values are better. Competing methods are described in~\cref{sec:synth-exper}. 
  }
  \label{fig:elpd}
\end{figure}

\begin{figure}[tp]
  \centering
  \includegraphics[width=\linewidth]{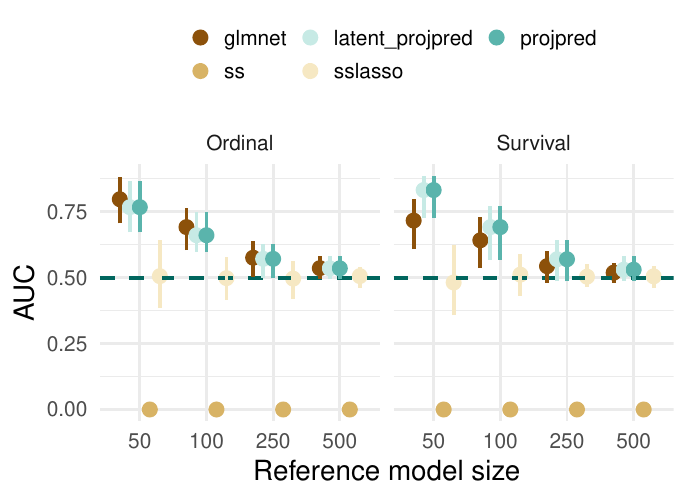}
  \caption{Median and 90\% confidence interval of the area under the curve (AUC) for the ROC for the selection of truly relevant terms in an ordinal cumulative model (left) and a time-to-event survival model (right) with $N= 100, \rho = 0$ for all competing methods. Higher values are better. Performance of pure random selection is shown as dashed dark green line. Competing methods are described in~\cref{sec:synth-exper}. 
  }
  \label{fig:optimal_size}
\end{figure}

We first fit ordinal and time-to-event survival regression models on extensive simulation conditions.
The generative process is
\begin{align*}
  x_n & \sim \mathrm{Normal}(0, \Sigma_{\rho}),\quad z_d  \sim \mathrm{Bernoulli}(0.6),\\
  \beta_d & \sim z_d\cdot\mathrm{Normal}(0, 1.5),\\
  \eta_n & \mid x_n, \beta = \beta^Tx_n,\quad y_n\mid \eta_n \sim p(g(\eta_n), \phi),
\end{align*}
where $y \in \mathbb{R}$, $x_n\in\mathrm{R}^D$ are the outcome and predictors respectively, $\beta$ denotes the unknown coefficients, number of variables $D$ is varied in $[50, 100, 250, 500]$ and the number of observations $N$ is also varied in $[100, 200, 300]$.
$\rho$ indicates the uniform correlation between predictors,  $g$ is the model-specific inverse link function and $p$ is either an ordinal cumulative likelihood (with $5$ ordered categories) or a time-to-event likelihood with Weibull base hazard process (maximum time-to-event is capped to $5$), and $\phi$ are model specific parameters.
Here, we show results for $N = 100$ observations and $\rho = 0$.
We show more simulations in the Appendix.

We fit the reference model making use of \texttt{brms}~\citep{B_rkner_2018} with a regularized horseshoe prior~\citep{piironen_sparsity_2017} accounting for the sparsity in the predictors.
The regularized horseshoe prior helps in avoiding overfitting, particularly for the higher dimensionality models.

Both the KL and residuals diagnostics indicate a good approximation for these models as shown in~\cref{sec:residuals-experiments}.
Here we analyse the predictive and selection performance.

\paragraph{Latent projections achieve the best held-out ELPD performance.}
The latent approach always achieves the best performance (\cref{fig:elpd}), even surpassing the full reference model with models that include substantially fewer terms.
This can be explained by a slight overfitting in the reference model.
In the most complex scenarios, we see that the performance of all methods suffers from slightly larger variance, including the full reference model.

\paragraph{Approximate likelihood methods result in underperforming selections (\cref{fig:elpd}).}
While the performance of \texttt{glmnet} is the closest to our proposed method on average, its performance distribution is wider and less reliable, while other methods suffer greatly from the approximate likelihood representation.
Spike-and-slab priors (\texttt{ss} in the figures) impose a very strong penalisation, which, together with the arbitrary choice of threshold ($0.5$ in our experiments, as is common, e.g. in~\cite{spikeslabgam}), can result in suboptimal selections.
The ill-informed approximate likelihood results in very few terms crossing the selection threshold, particularly in \texttt{ss} and naive \texttt{projpred}.
Even though \texttt{sslasso} does not need an arbitrary threshold for the selection, it suffers from similar problems. 

\paragraph{Other approaches require refitting a Bayesian model.}
Other approaches operate on an approximate likelihood model that does not result in a sensible predictive model.
To have a functional reduced model in the same domain as the reference model, all approaches except ours require refitting a full Bayesian model with the selected predictors and correct (non-normal) likelihood, increasing their cost.
Our latent approach fully honours the reference model structure, and predictions in the original model space only require evaluating the inverse link function for the latent predictions. 

\paragraph{Latent approach retains reference model uncertainty.}
The uncertainty present in the reference model is projected as well, contributing to better informed solutions than the other approaches, which cannot represent the original model likelihood.
As seen in \cref{fig:optimal_size}, both \texttt{projpred} approaches achieve the highest AUC, even though naive \texttt{projpred} fails to assess the predictive performance of the projections.
For the simpler \texttt{ordinal} experiments, \texttt{glmnet} achieves similar AUC.
Interestingly, \texttt{ss} has trouble identifying any relevant features, and therefore its selection is not accurate, whereas \texttt{sslasso} manages to identify at least some.

\subsection{Non-exponential family models with real data}
\label{sec:real-data-sets}
In this section, we assess the selection and predictive capabilities of our latent approach on real datasets and compare it against \texttt{glmnet} as the best competitor in our simulated examples.
We focus on the quality of the selection path and the properties of the latent projections.
A brief description of each dataset can be found in the Appendix.

\begin{figure*}
	\centering
	\begin{subfigure}{0.245\linewidth}
    \includegraphics[height=90pt, width=\linewidth]{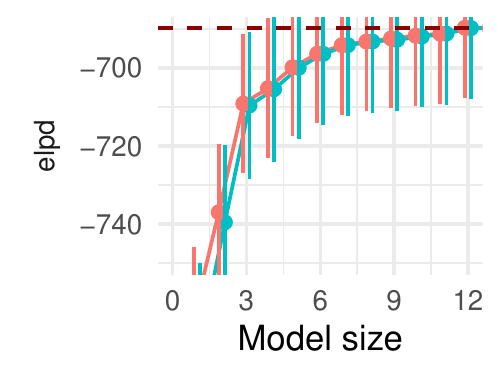}
    \caption{Eyedisease}
	\end{subfigure}
	\begin{subfigure}{0.245\linewidth}
    \includegraphics[height=90pt, width=\linewidth]{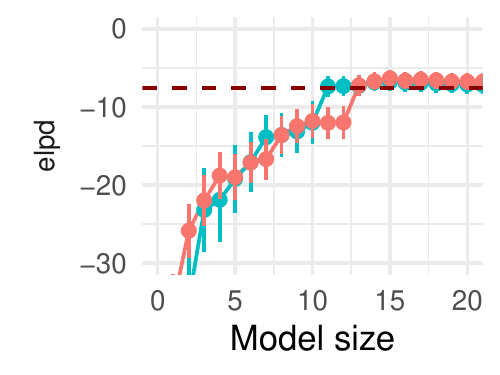}
    \caption{HCCFrame}
	\end{subfigure}
	\begin{subfigure}{0.245\linewidth}
    \includegraphics[height=90pt, width=\linewidth]{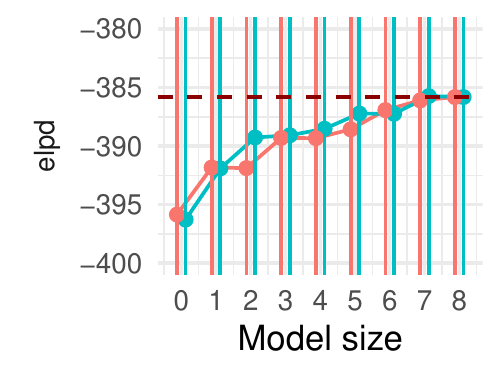}
    \caption{Cancer}
	\end{subfigure}
	\begin{subfigure}{0.245\linewidth}
    \includegraphics[height=90pt, width=\linewidth]{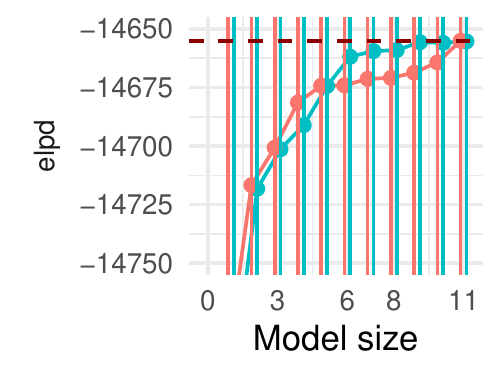}
    \caption{Rotterdam}
	\end{subfigure}
	\caption{Full data zoomed-in ELPD performance comparison in the response model space. For \texttt{latent\_projpred} the projections are computed in the latent space and the predictions are then transformed. Red shows \texttt{glmnet} and blue \texttt{latent\_projpred}. Error bars indicate ELPD 95\% quantiles. Larger error bars correspond to datasets with fewer observations.}
	\label{fig:real_datasets_path_comparison}
\end{figure*}

\begin{figure*}[tp]
	\centering
	\begin{subfigure}{0.245\linewidth}
    \includegraphics[height=90pt, width=\linewidth]{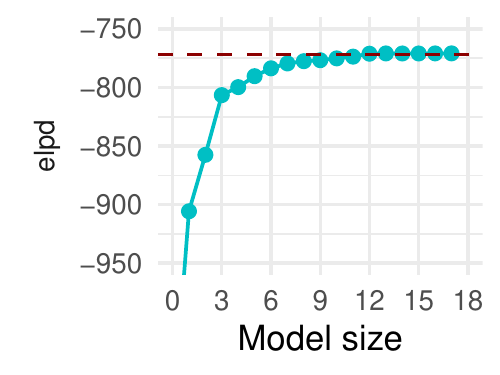}
    \caption{Eyedisease}
    \label{fig:eyedisease_latent}	
	\end{subfigure}
	\begin{subfigure}{0.245\linewidth}
    \includegraphics[height=90pt, width=\linewidth]{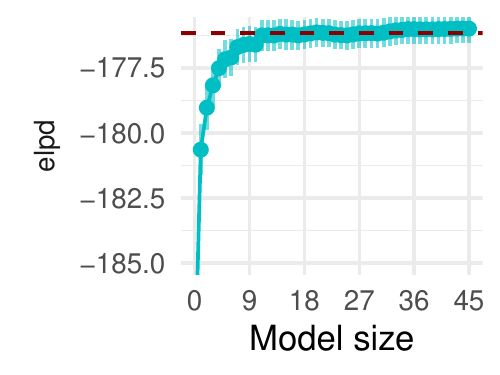}
    \caption{HCCFrame}
    \label{fig:hccframe_latent}
	\end{subfigure}
	\begin{subfigure}{0.245\linewidth}
    \includegraphics[height=90pt, width=\linewidth]{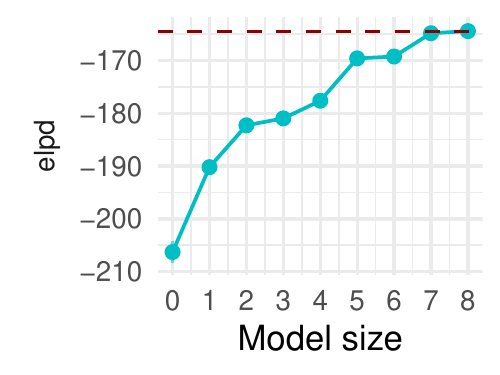}
    \caption{Cancer}
    \label{fig:eyedisease_latent}
	\end{subfigure}
	\begin{subfigure}{0.245\linewidth}
    \includegraphics[height=90pt, width=\linewidth]{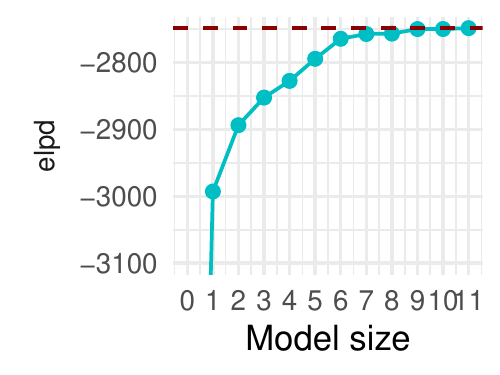}
    \caption{Rotterdam}
    \label{fig:hccframe_latent}
	\end{subfigure}
	\caption{Full data ELPD performance in the latent predictor space for \texttt{latent\_projpred}. Reference performance in dark red dashed horizontal line.}
	\label{fig:real_datasets_projpred}
\end{figure*}

\paragraph{Latent projections identify superior solution path.}
We show the comparison of the latent approach against \texttt{glmnet} in~\cref{fig:real_datasets_path_comparison}.
From a predictive performance point of view, the latent approach identifies a better solution path for all datasets we tested, except for \texttt{eyedisease}, where both approaches are equal.
Note that, along the path, there are multiple model sizes for which the latent approach's projections' performance is superior to their \texttt{glmnet} counterpart.
Datasets with few observations (\texttt{cancer} and \texttt{rotterdam} particularly) suffer from higher variance in performance.

\paragraph{The response space is noisier.}
Our latent approach allows us to perform the variable selection on either the response scale or the latent predictor scale. 
When compared to the selection in response scale (as seen in~\cref{fig:real_datasets_path_comparison}), the latent predictor space is significantly less noisy (\cref{fig:real_datasets_projpred}), mostly due to the complex observation model and sometimes relatively small datasets.
The reference model filters noise from the data and thus the latent variables have less noise, resulting in reduced variance in the selection criterion (\cref{fig:real_datasets_projpred} \emph{Latent} label), even in data with few observations (\texttt{Cancer} and \texttt{Rotterdam} in particular).

\begin{figure}[tp]
    \centering
    \includegraphics[width=\linewidth]{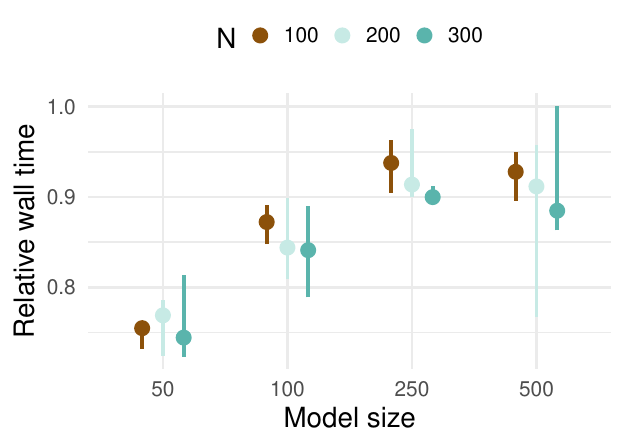}
    \caption{Median and $90$\% confidence interval relative wall time of \texttt{latent\_projpred} over \texttt{projpred} for $50$ Bernoulli data realizations and varying number of observations (different shapes). Lower than $1$ means \texttt{latent\_projpred} is faster.}
    \label{fig:time_comparison}
\end{figure}

\begin{figure}[tp]
    \centering
    \includegraphics[height=185pt,width=\linewidth]{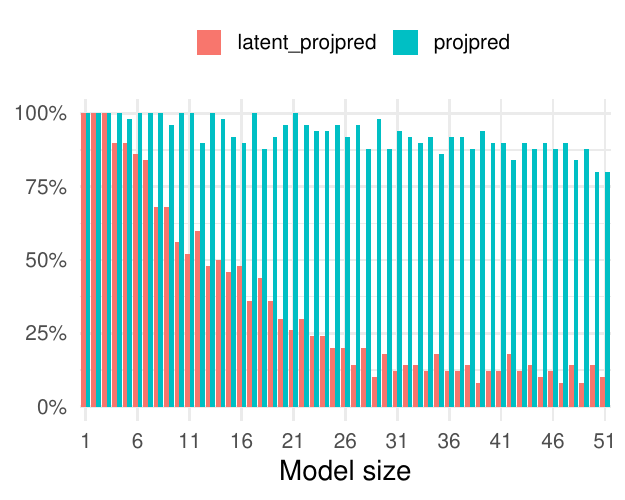}
    \caption{Bootstrap inclusion frequencies for a simulated Bernoulli dataset with $N = 100$ observations and $D = 50$ variables computed from $50$ boostrap samples.}
    \label{fig:bootstrap_comparison}
\end{figure}

\subsection{Non-Gaussian exponential family models with simulated data}
\label{sec:latent-benefits}

We simulate $50$ realizations of a Bernoulli data with $N = 300$ observations and $D = 50, 100, 250, 500$ uncorrelated variables.

\paragraph{The latent approach is faster to compute.} 
The latent normal approximation is solved efficiently in closed form, which is faster to compute than the iterative solution of the original projection approach, resulting in significantly faster solutions (\cref{fig:time_comparison}).

\paragraph{Latent approach has less variability and selects fewer variables.} 
We simulate $50$ bootstrap samples from the same Bernoulli data generation process with $N = 100$ observations and $D = 50$ variables and compare the selection of the original framework against the latent approach.
As seen in~\cref{fig:bootstrap_comparison}, the latent approach results in smaller subsets of relevant variables with a smaller variability across bootstrap samples, while the original framework fails at identifying a suitable subset and overselects variables in every case. 

\begin{figure}[tp]
  \centering
  \includegraphics[height=180pt,width=\linewidth]{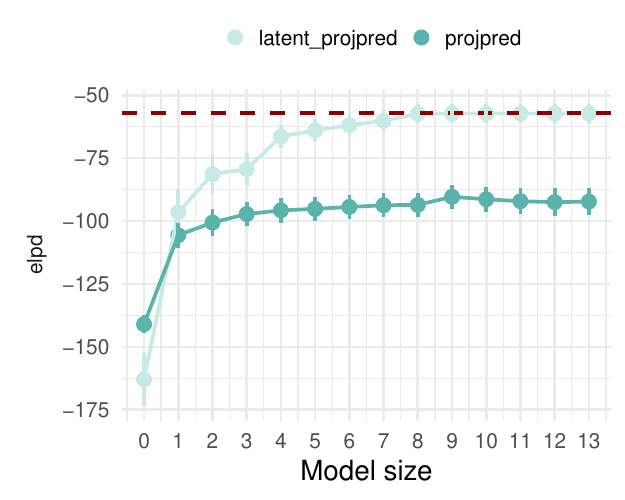}
  \caption{Comparison of full data expected log predictive density (ELPD) performance for a hierarchical Bernoulli model of fertility among \kakapo population. 
  Reference performance in dark red dashed horizontal line. The original \texttt{projpred} solution does not reach the reference model predictive performance due to convergence issues in the underlying projections when including the varying intercept per individual. 
  }
  \label{fig:kakapo_latent_vs_bernoulli}
\end{figure}

\subsection{Non-Gaussian exponential family models with real data}
\label{sec:kakapo-latent}

For complex non-Gaussian models, the original framework provides a solution that might suffer from unstable projections and often fail to converge to sensible results.
We show that the latent approach obtains a superior solution for a real model in the context of endangered species conservation taken from an ongoing collaboration with domain experts~\citep{kakapo}. 

These data consist of a set of 211 \kakapo individuals (it is not possible to obtain more observations due to the very nature of the data set) and the aim of the study is to find a model for the fertility of upcoming eggs.
With the domain experts, we constructed a hierarchical Bernoulli model (the response is either 0 or 1 for fertile and infertile eggs) with a varying intercept per individual.
Such group effects can soak most of the outcome variance in the model, which paired with the nonlinear link function, can often result in PIRLS solver convergence issues.

\paragraph{Latent approach provides superior performing projections (\cref{fig:kakapo_latent_vs_bernoulli}).}
Since the latent predictor is naturally continuous and unbounded, the latent approach results in more stable and efficient projections for non-Gaussian exponential family models, as seen in~\cref{fig:kakapo_latent_vs_bernoulli}.
The latent approach accurately captures the reference model predictive performance, and also a better ordering in the solution path.

\section{Discussion}
\label{sec:org755be9f}

We proposed a novel latent space projection predictive approach to perform variable and structure selection in models with non-exponential family observation model.
As shown in the experiments, the approach is also beneficial for non-Gaussian exponential family models.

We have shown that our approach offers superior performance in extensive experiments on high dimensional generalized linear models and several real datasets. 
The proposed method is not limited to latent linear models, as it can readily be used for variable and structure selection in more complicated models, including linear and non-linear hierarchical models (as demonstrated in~\cref{fig:kakapo_latent_vs_bernoulli}).

Our approach not only improves variable and structure selection in a whole new set of models while respecting their original structure, but also offers superior performance in currently supported models in \texttt{projpred}. 

The method may have worse performance for models whose latent space is far from normal. 
We provide diagnostics that alert the user in these cases, but further research is needed to improve the performance in such cases.

\subsubsection*{Acknowledgments}
The authors thank Aalto Science-IT for computational resources and FCAI for funding and support.

\bibliographystyle{abbrvnat}
\bibliography{biblio}

\begin{thebibliography}{44}
\providecommand{\natexlab}[1]{#1}
\providecommand{\url}[1]{\texttt{#1}}
\expandafter\ifx\csname urlstyle\endcsname\relax
  \providecommand{\doi}[1]{doi: #1}\else
  \providecommand{\doi}{doi: \begingroup \urlstyle{rm}\Url}\fi

\bibitem[Archer and Williams(2012)]{glmnetcr}
K.~J. Archer and A.~A. Williams.
\newblock {L1} penalized continuation ratio models for ordinal response
  prediction using high-dimensional datasets.
\newblock \emph{Statistics in Medicine}, 31:\penalty0 1464--1474, 2012.
\newblock URL \url{https://www.ncbi.nlm.nih.gov/pmc/articles/PMC3718008/}.

\bibitem[Archer et~al.(2014)Archer, Hou, Zhou, Ferber, Layne, and
  Gentry]{ordinalgmifs}
K.~J. Archer, J.~Hou, Q.~Zhou, K.~Ferber, J.~G. Layne, and A.~E. Gentry.
\newblock ordinalgmifs: An {R} package for ordinal regression in
  high-dimensional data settings.
\newblock \emph{Cancer informatics}, 13:\penalty0 187—195, 2014.
\newblock ISSN 1176-9351.
\newblock \doi{10.4137/cin.s20806}.
\newblock URL \url{https://europepmc.org/articles/PMC4266195}.

\bibitem[Barron(1992)]{barron1992analysis}
D.~N. Barron.
\newblock The analysis of count data: Overdispersion and autocorrelation.
\newblock \emph{Sociological methodology}, pages 179--220, 1992.

\bibitem[Blei et~al.(2017)Blei, Kucukelbir, and McAuliffe]{blei2017variational}
D.~M. Blei, A.~Kucukelbir, and J.~D. McAuliffe.
\newblock Variational inference: A review for statisticians.
\newblock \emph{Journal of the American statistical Association}, 112\penalty0
  (518):\penalty0 859--877, 2017.

\bibitem[Breiman(1995)]{breiman_garrote_1995}
L.~Breiman.
\newblock Better subset regression using the nonnegative garrote.
\newblock \emph{Technometrics}, 37\penalty0 (4):\penalty0 373--384, 1995.
\newblock ISSN 00401706.
\newblock URL \url{http://www.jstor.org/stable/1269730}.

\bibitem[B\"urkner(2018)]{B_rkner_2018}
P.-C. B\"urkner.
\newblock Advanced {Bayesian} multilevel modeling with the r package brms.
\newblock \emph{The R Journal}, 10\penalty0 (1):\penalty0 395, 2018.
\newblock ISSN 2073-4859.
\newblock \doi{10.32614/rj-2018-017}.
\newblock URL \url{http://dx.doi.org/10.32614/rj-2018-017}.

\bibitem[B{\"u}rkner and Vuorre(2019)]{burkner2019ordinal}
P.-C. B{\"u}rkner and M.~Vuorre.
\newblock Ordinal regression models in psychology: A tutorial.
\newblock \emph{Advances in Methods and Practices in Psychological Science},
  2\penalty0 (1):\penalty0 77--101, 2019.

\bibitem[Candes and Tao(2007)]{Candes_2007}
E.~Candes and T.~Tao.
\newblock The dantzig selector: Statistical estimation when p is much larger
  than n.
\newblock \emph{The Annals of Statistics}, 35\penalty0 (6):\penalty0
  2313–2351, Dec 2007.
\newblock ISSN 0090-5364.
\newblock \doi{10.1214/009053606000001523}.
\newblock URL \url{http://dx.doi.org/10.1214/009053606000001523}.

\bibitem[Carvalho et~al.(2010)Carvalho, Polson, and Scott]{Carvalho_2010}
C.~M. Carvalho, N.~G. Polson, and J.~G. Scott.
\newblock The horseshoe estimator for sparse signals.
\newblock \emph{Biometrika}, 97\penalty0 (2):\penalty0 465–480, Apr 2010.
\newblock ISSN 1464-3510.
\newblock \doi{10.1093/biomet/asq017}.
\newblock URL \url{http://dx.doi.org/10.1093/biomet/asq017}.

\bibitem[Catalina et~al.(2020)Catalina, Bürkner, and
  Vehtari]{catalina2020projection}
A.~Catalina, P.-C. Bürkner, and A.~Vehtari.
\newblock Projection predictive inference for generalized linear and additive
  multilevel models.
\newblock \emph{arxiv preprint:2010.06994}, 2020.

\bibitem[Dhaka et~al.(2020)Dhaka, Catalina, Andersen, Magnusson, Huggins, and
  Vehtari]{DBLP:conf/nips/DhakaCAMHV20}
A.~K. Dhaka, A.~Catalina, M.~R. Andersen, M.~Magnusson, J.~H. Huggins, and
  A.~Vehtari.
\newblock Robust, accurate stochastic optimization for variational inference.
\newblock In \emph{NeurIPS}, 2020.
\newblock URL
  \url{https://proceedings.neurips.cc/paper/2020/hash/7cac11e2f46ed46c339ec3d569853759-Abstract.html}.

\bibitem[Digby et~al.(2021)Digby, Eason, Catalina, Programme, Lierz, Galla,
  authors TBC, Steeves, and Vercoe]{kakapo}
A.~Digby, D.~Eason, A.~Catalina, K.~R. Programme, M.~Lierz, S.~Galla, G.~A.
  authors TBC, T.~Steeves, and D.~Vercoe.
\newblock Captive rearing reduces fertility in the critically endangered
  kakapo, 2021.

\bibitem[Dupuis and Robert(2003)]{Dupuis2003}
J.~A. Dupuis and C.~P. Robert.
\newblock Variable selection in qualitative models via an entropic explanatory
  power.
\newblock \emph{Journal of Statistical Planning and Inference}, 111, 2003.
\newblock ISSN 03783758.
\newblock \doi{10.1016/S0378-3758(02)00286-0}.

\bibitem[Fan and Li(2001)]{fan_nonconcave_2001}
J.~Fan and R.~Li.
\newblock Variable selection via nonconcave penalized likelihood and its oracle
  properties.
\newblock \emph{Journal of the American Statistical Association}, 96\penalty0
  (456):\penalty0 1348--1360, 2001.
\newblock \doi{10.1198/016214501753382273}.
\newblock URL \url{https://doi.org/10.1198/016214501753382273}.

\bibitem[Fan et~al.(2005)Fan, Li, and Li]{fan2005overview}
J.~Fan, G.~Li, and R.~Li.
\newblock An overview on variable selection for survival analysis.
\newblock In \emph{Contemporary Multivariate Analysis And Design Of
  Experiments: In Celebration of Professor Kai-Tai Fang's 65th Birthday}, pages
  315--336. World Scientific, 2005.

\bibitem[Friedman et~al.(2010{\natexlab{a}})Friedman, Hastie, and
  Tibshirani]{Friedman_2010}
J.~Friedman, T.~Hastie, and R.~Tibshirani.
\newblock Regularization paths for generalized linear models via coordinate
  descent.
\newblock \emph{Journal of Statistical Software}, 33\penalty0 (1),
  2010{\natexlab{a}}.
\newblock ISSN 1548-7660.
\newblock \doi{10.18637/jss.v033.i01}.
\newblock URL \url{http://dx.doi.org/10.18637/jss.v033.i01}.

\bibitem[Friedman et~al.(2010{\natexlab{b}})Friedman, Hastie, and
  Tibshirani]{glmnet}
J.~Friedman, T.~Hastie, and R.~Tibshirani.
\newblock Regularization paths for generalized linear models via coordinate
  descent.
\newblock \emph{Journal of Statistical Software}, 33\penalty0 (1):\penalty0
  1--22, 2010{\natexlab{b}}.
\newblock URL \url{http://www.jstatsoft.org/v33/i01/}.

\bibitem[Gelman et~al.(2020)Gelman, Vehtari, Simpson, Margossian, Carpenter,
  Yao, Kennedy, Gabry, Bürkner, and Modrák]{gelman2020bayesian}
A.~Gelman, A.~Vehtari, D.~Simpson, C.~C. Margossian, B.~Carpenter, Y.~Yao,
  L.~Kennedy, J.~Gabry, P.-C. Bürkner, and M.~Modrák.
\newblock Bayesian workflow, 2020.

\bibitem[Goutis and Robert(1998)]{Goutis_1998}
C.~Goutis and C.~P. Robert.
\newblock Model choice in generalised linear models: {A} {Bayesian} approach
  via {Kullback-Leibler} projections.
\newblock \emph{Biometrika}, 85\penalty0 (1):\penalty0 29–37, Mar 1998.
\newblock ISSN 1464-3510.
\newblock \doi{10.1093/biomet/85.1.29}.
\newblock URL \url{http://dx.doi.org/10.1093/biomet/85.1.29}.

\bibitem[Hastie(2015)]{Hastie_2015}
T.~Hastie.
\newblock \emph{Statistical Learning with Sparsity}.
\newblock Chapman and Hall/CRC, May 2015.
\newblock ISBN 9781498712170.
\newblock \doi{10.1201/b18401}.
\newblock URL \url{http://dx.doi.org/10.1201/b18401}.

\bibitem[Huszár(2017)]{DBLP:journals/corr/Huszar17}
F.~Huszár.
\newblock Variational inference using implicit distributions.
\newblock \emph{CoRR}, abs/1702.08235, 2017.
\newblock URL \url{http://arxiv.org/abs/1702.08235}.

\bibitem[Ishwaran and Rao(2005)]{Ishwaran_2005}
H.~Ishwaran and J.~S. Rao.
\newblock Spike and slab variable selection: Frequentist and {Bayesian}
  strategies.
\newblock \emph{The Annals of Statistics}, 33\penalty0 (2):\penalty0 730–773,
  Apr 2005.
\newblock ISSN 0090-5364.
\newblock \doi{10.1214/009053604000001147}.
\newblock URL \url{http://dx.doi.org/10.1214/009053604000001147}.

\bibitem[Kelter(2020)]{kelter_survival}
R.~Kelter.
\newblock Bayesian survival analysis in stan for improved measuring of
  uncertainty in parameter estimates.
\newblock \emph{Measurement: Interdisciplinary Research and Perspectives},
  18\penalty0 (2):\penalty0 101--109, 2020.
\newblock \doi{10.1080/15366367.2019.1689761}.
\newblock URL \url{https://doi.org/10.1080/15366367.2019.1689761}.

\bibitem[Koopman(1936)]{Koopman_1936}
B.~O. Koopman.
\newblock On distributions admitting a sufficient statistic.
\newblock \emph{Transactions of the American Mathematical Society}, 39\penalty0
  (3):\penalty0 399–399, Mar 1936.
\newblock ISSN 0002-9947.
\newblock \doi{10.1090/s0002-9947-1936-1501854-3}.
\newblock URL \url{http://dx.doi.org/10.1090/s0002-9947-1936-1501854-3}.

\bibitem[Marx(1996)]{marx1996iteratively}
B.~D. Marx.
\newblock Iteratively reweighted partial least squares estimation for
  generalized linear regression.
\newblock \emph{Technometrics}, 38\penalty0 (4):\penalty0 374--381, 1996.

\bibitem[McCullagh and Nelder(1989)]{McCullagh_1989}
P.~McCullagh and J.~A. Nelder.
\newblock \emph{Generalized Linear Models}.
\newblock Springer US, 1989.
\newblock ISBN 9781489932426.
\newblock \doi{10.1007/978-1-4899-3242-6}.
\newblock URL \url{http://dx.doi.org/10.1007/978-1-4899-3242-6}.

\bibitem[Nagler(1994)]{nagler_ordinal}
J.~Nagler.
\newblock Scobit: An alternative estimator to logit and probit.
\newblock \emph{American Journal of Political Science}, 38\penalty0
  (1):\penalty0 230--255, 1994.
\newblock ISSN 00925853, 15405907.
\newblock URL \url{http://www.jstor.org/stable/2111343}.

\bibitem[O'Hara and Sillanpää(2009)]{ohara2009review}
R.~B. O'Hara and M.~J. Sillanpää.
\newblock {A review of Bayesian variable selection methods: what, how and
  which}.
\newblock \emph{Bayesian Analysis}, 4\penalty0 (1):\penalty0 85 -- 117, 2009.
\newblock \doi{10.1214/09-BA403}.
\newblock URL \url{https://doi.org/10.1214/09-BA403}.

\bibitem[Pavone et~al.(2020)Pavone, Piironen, B{\"u}rkner, and
  Vehtari]{pavone_2019}
F.~Pavone, J.~Piironen, P.-C. B{\"u}rkner, and A.~Vehtari.
\newblock Using reference models in variable selection.
\newblock \emph{arXiv preprint arXiv:2004.13118}, 2020.

\bibitem[Pequignot et~al.(2020)Pequignot, Alain, Dallaire, Yeganehparast,
  Germain, Desharnais, and Laviolette]{pequignot2020implicit}
Y.~Pequignot, M.~Alain, P.~Dallaire, A.~Yeganehparast, P.~Germain,
  J.~Desharnais, and F.~Laviolette.
\newblock Implicit variational inference: the parameter and the predictor
  space, 2020.

\bibitem[Piironen and Vehtari(2016)]{piironen_projection_2016}
J.~Piironen and A.~Vehtari.
\newblock Projection predictive model selection for {Gaussian} processes.
\newblock In \emph{2016 {IEEE} 26th International Workshop on Machine Learning
  for Signal Processing ({MLSP})}, pages 1--6, 2016.
\newblock \doi{10.1109/MLSP.2016.7738829}.

\bibitem[Piironen and Vehtari(2017{\natexlab{a}})]{piironen_comparison_2017}
J.~Piironen and A.~Vehtari.
\newblock Comparison of {Bayesian} predictive methods for model selection.
\newblock \emph{Statistics and Computing}, 27\penalty0 (3):\penalty0 711--735,
  2017{\natexlab{a}}.
\newblock ISSN 0960-3174, 1573-1375.
\newblock \doi{10.1007/s11222-016-9649-y}.
\newblock URL \url{http://link.springer.com/10.1007/s11222-016-9649-y}.

\bibitem[Piironen and Vehtari(2017{\natexlab{b}})]{piironen_sparsity_2017}
J.~Piironen and A.~Vehtari.
\newblock Sparsity information and regularization in the horseshoe and other
  shrinkage priors.
\newblock \emph{Electronic Journal of Statistics}, 11\penalty0 (2):\penalty0
  5018--5051, 2017{\natexlab{b}}.
\newblock ISSN 1935-7524.
\newblock \doi{10.1214/17-EJS1337SI}.
\newblock URL \url{https://projecteuclid.org/euclid.ejs/1513306866}.

\bibitem[Piironen et~al.(2020{\natexlab{a}})Piironen, Paasiniemi, Catalina, and
  Vehtari]{projpred_package}
J.~Piironen, M.~Paasiniemi, A.~Catalina, and A.~Vehtari.
\newblock \emph{projpred: Projection Predictive Feature Selection},
  2020{\natexlab{a}}.
\newblock https://mc-stan.org/projpred, https://discourse.mc-stan.org/.

\bibitem[Piironen et~al.(2020{\natexlab{b}})Piironen, Paasiniemi, and
  Vehtari]{piironen_projective_2018}
J.~Piironen, M.~Paasiniemi, and A.~Vehtari.
\newblock Projective inference in high-dimensional problems: Prediction and
  feature selection.
\newblock \emph{Electronic Journal of Statistics}, 14\penalty0 (1):\penalty0
  2155--2197, 2020{\natexlab{b}}.
\newblock \doi{10.1214/20-EJS1711}.
\newblock URL \url{https://doi.org/10.1214/20-EJS1711}.

\bibitem[Rezende and Mohamed(2016)]{rezende2016variational}
D.~J. Rezende and S.~Mohamed.
\newblock Variational inference with normalizing flows, 2016.

\bibitem[Robert and Casella(2013)]{robert2013monte}
C.~Robert and G.~Casella.
\newblock \emph{Monte Carlo statistical methods}.
\newblock Springer Science \& Business Media, 2013.

\bibitem[Rockova and George(2018)]{sslasso}
V.~Rockova and E.~George.
\newblock The spike-and-slab lasso.
\newblock \emph{Journal of the American Statistical Association}, 113\penalty0
  (521):\penalty0 431--444, 2018.

\bibitem[Scheipl(2011)]{spikeslabgam}
F.~Scheipl.
\newblock {spikeSlabGAM}: {B}ayesian variable selection, model choice and
  regularization for generalized additive mixed models in {R}.
\newblock \emph{Journal of Statistical Software}, 43\penalty0 (14):\penalty0
  1--24, 2011.
\newblock URL \url{http://www.jstatsoft.org/v43/i14/}.

\bibitem[Tibshirani(1996)]{tibshirani_lasso_1996}
R.~Tibshirani.
\newblock Regression shrinkage and selection via the {Lasso}.
\newblock \emph{Journal of the Royal Statistical Society. Series B
  (Methodological)}, 58\penalty0 (1):\penalty0 267--288, 1996.
\newblock ISSN 00359246.
\newblock URL \url{http://www.jstor.org/stable/2346178}.

\bibitem[Titsias and Ruiz(2019)]{titsias2019unbiased}
M.~K. Titsias and F.~Ruiz.
\newblock Unbiased implicit variational inference.
\newblock In \emph{The 22nd International Conference on Artificial Intelligence
  and Statistics}, pages 167--176. PMLR, 2019.

\bibitem[Vehtari and Ojanen(2012)]{aki_model_assessment}
A.~Vehtari and J.~Ojanen.
\newblock {A survey of Bayesian predictive methods for model assessment,
  selection and comparison}.
\newblock \emph{Statistics Surveys}, 6\penalty0 (none):\penalty0 142 -- 228,
  2012.
\newblock \doi{10.1214/12-SS102}.
\newblock URL \url{https://doi.org/10.1214/12-SS102}.

\bibitem[Wurm et~al.(2017)Wurm, Rathouz, and
  Hanlon]{wurm17:_regul_ordin_regres_r_packag}
M.~J. Wurm, P.~J. Rathouz, and B.~M. Hanlon.
\newblock {Regularized Ordinal Regression and the ordinalNet R Package}, 2017.

\bibitem[Zou and Hastie(2005)]{Zou_2005}
H.~Zou and T.~Hastie.
\newblock Regularization and variable selection via the elastic net.
\newblock \emph{Journal of the Royal Statistical Society: Series B (Statistical
  Methodology)}, 67\penalty0 (2):\penalty0 301–320, Apr 2005.
\newblock ISSN 1467-9868.
\newblock \doi{10.1111/j.1467-9868.2005.00503.x}.
\newblock URL \url{http://dx.doi.org/10.1111/j.1467-9868.2005.00503.x}.

\end{thebibliography}
\end{document}